\pdfoutput=0
\documentclass[twocolumn,showpacs,pre,preprintnumbers,amsmath,amssymb,aps,floatfix]{revtex4}
\usepackage{graphicx}
\usepackage{dcolumn}
\usepackage{bm}
\usepackage{subfigure}
\begin{document}
\title{Increased accuracy  of  ligand sensing by  receptor diffusion on cell surface}
\author{Gerardo Aquino}
\author{Robert G.  Endres }
\affiliation{Division of Molecular Biosciences and Centre for Integrated Systems Biology at Imperial College, Imperial College London,  SW7 2AZ, London, UK}
\date{\today }

\newcommand{\Sss}{\scriptscriptstyle}
\newcommand{\Ss}{\scriptstyle}
\newcommand{\D}{\dysplaystyle}
\newcommand{\T}{\textstyle}
\newcommand{\e}{{\rm e}}
\newcommand{\veps}{\varepsilon}
\newcommand{\epss}{\varepsilon_{\sigma}}
\newcommand{\epsv}{\varepsilon_{V}}
\newcommand{\lgl}{\langle}
 \newcommand{\rgl}{\rangle}
\newcommand{\Vh}[1]{\hat{#1}}
\newcommand{\Aa}{A^1_{\epsilon}}
\newcommand{\Ab}{A^{\epsilon}_L}
\newcommand{\Ae}{A_{\epsilon}}
\newcommand{\finn}[1]{\phi^{\pm}_{#1}}
\newcommand{\ea}{e^{-|\alpha|^2}}
\newcommand{\eb}{\frac{e^{-|\alpha|^2} |\alpha|^{2 n}}{n!}}
\newcommand{\ebbb}{\frac{e^{-3|\alpha|^2} |\alpha|^{2 (l+n+m)}}{l!m!n!}}
\newcommand{\ass}{\alpha}
\newcommand{\as}{\alpha^*}
\newcommand{\fb}{\bar{f}}
\newcommand{\gb}{\bar{g}}
\newcommand{\la}{\lambda}
\newcommand{\sz}{\hat{s}_{z}}
\newcommand{\sy}{\hat{s}_y}
\newcommand{\sx}{\hat{s}_x}
\newcommand{\sio}{\hat{\sigma}_0}
\newcommand{\six}{\hat{\sigma}_x}
\newcommand{\siz}{\hat{\sigma}_{z}}
\newcommand{\siy}{\hat{\sigma}_y}
\newcommand{\vhsig}{\vec{\hat{\sigma}}}
\newcommand{\hsig}{\hat{\sigma}}
\newcommand{\hH}{\hat{H}}
\newcommand{\hU}{\hat{U}}
\newcommand{\hA}{\hat{A}}
\newcommand{\hB}{\hat{B}}
\newcommand{\hC}{\hat{C}}
\newcommand{\hD}{\hat{D}}
\newcommand{\hV}{\hat{V}}
\newcommand{\hW}{\hat{W}}
\newcommand{\hK}{\hat{K}}
\newcommand{\hX}{\hat{X}}
\newcommand{\hM}{\hat{M}}
\newcommand{\hN}{\hat{N}}
\newcommand{\kf}{k_i^e}
\newcommand{\te}{\theta}
\newcommand{\vze}{\vec{\zeta}}
\newcommand{\vet}{\vec{\eta}}
\newcommand{\vx}{\vec{\xi}}
\newcommand{\vc}{\vec{\chi}}
\newcommand{\hro}{\hat{\rho}_b}
\newcommand{\vro}{\vec{\rho}_b}
\newcommand{\hR}{\hat{R}}
\newcommand{\half}{\frac{1}{2}}
\renewcommand{\d}{{\rm d}}
\renewcommand{\top }{ t^{\prime } }
\newcommand{\oz}{{(0)}}
\newcommand{\sint}{{\rm si}}
\newcommand{\cint}{{\rm ci}}
\newcommand{\de}{\delta}
\newcommand{\ep}{\varepsilon}
\newcommand{\De}{\Delta}
\newcommand{\eps}{\varepsilon}
\newcommand{\si}{\hat{\sigma}}
\newcommand{\om}{\omega}
\newcommand{\tr}{{\rm tr}}
\newcommand{\ha}{\hat{a}}
\newcommand{\gam}{\gamma ^{(0)}}
\newcommand{\pe}{\prime}
\newcommand{\BEQ}{\begin{equation}}
\newcommand{\EEQ}{\end{equation}}
\newcommand{\BEGN}{\begin{align}}
\newcommand{\EEGN}{\end{align}}
\newcommand{\BEFGN}{\begin{flalign}}
\newcommand{\EEFGN}{\end{flalign}}
\newcommand{\BES}{\begin{subequations}}
\newcommand{\EES}{\end{subequations}}
\newcommand{\BEA}{\begin{eqnarray}}
\newcommand{\EEA}{\end{eqnarray}}
\newcommand{\sph}{spin-$\frac{1}{2}$ }
\newcommand{\ad}{\hat{a}^{\dagger}}
\newcommand{\add}{\hat{a}}
\newcommand{\spp}{\hat{\sigma}_+}
\newcommand{\smm}{\hat{\sigma}_-}
\newcommand{\fin}[1]{|\phi^{\pm}_{#1}\rangle}
\newcommand{\finm}[1]{|\phi^{-}_{#1}\rangle}
\newcommand{\lfin}[1]{\langle \phi^{\pm}_{#1}|}
\newcommand{\lfinp}[1]{\langle \phi^{+}_{#1}|}
\newcommand{\lfinm}[1]{\langle \phi^{-}_{#1}|}
\newcommand{\lfinn}[1]{\langle\phi^{\pm}_{#1}|}
\newcommand{\z}{\cal{Z}}
\newcommand{\RI}{\hat{{\cal{R}}}_{0}}
\newcommand{\Rt}{\hat{{\cal{R}}}_{\tau}}
\newcommand{\cb}{\bar{c}}
\newcommand{\nb}{\bar{n}}
\newcommand{\dnr}{ \delta n(\vec{r},t)}
\newcommand{\dnz}{ \delta n(\vec{r}_0,t)}
\newcommand{\dn}{ \delta n(t)}
\newcommand{\km}{ \kappa_{-}}
\newcommand{\dc}{ \delta c(\vec{x},t)}
\newcommand{\dcz}{ \delta c(\vec{x}_0,t)}
\newcommand{\dcw}{ \delta \hat{c}(\vec{x}_0,\omega)}
\newcommand{\dch}{ \delta \hat{c}(\vec{q},\omega)}
\newcommand{\dxi}{ \hat{\xi}_c(\vec{q},\omega)}
\newcommand{\dxib}{ \hat{\xi}_c(\omega)}
\newcommand{\dnh}{ \delta \hat{n}(\omega)}
\newcommand{\dnhq}{ \delta \hat{n}(\vec{q},\omega)}
\newcommand{\dnhz}{ \delta \hat{n}(\vec{r}_0,\omega)}
\newcommand{\dchz}{ \delta \hat{c}(\vec{r}_0,z_0,\omega)}
\newcommand{\nv}{  n(\vec{r},t)}
\newcommand{\cv}{ c(\vec{r},t)}
\newcommand{\nn}{\nonumber}
\newcommand{\rnb}{(\rho_0-\nb)}
\newcommand{\rnbo}{(1-\nb)}

\begin{abstract}
The physical limit with which a cell senses external ligand concentration
corresponds to the perfect absorber,  where all ligand particles are absorbed and overcounting of same ligand particles does not occur.
Here we analyze how the lateral diffusion of receptors on the cell membrane 
 affects the accuracy of sensing  ligand concentration.
Specifically, we connect our modeling to 
 neurotransmission in neural synapses
where the diffusion of glutamate receptors is  already known to 
 refresh synaptic connections.
We find that  receptor diffusion indeed  increases the accuracy of sensing  
for both the glutamate AMPA  and
 NDMA receptors, although the NMDA receptor is overall much noiser. We propose that the difference in accuracy of sensing of the two receptors can 
be linked to their different roles
  in neurotransmission. Specifically, the high accuracy in sensing 
glutamate is essential for the AMPA receptor to start membrane depolarization, while
the NMDA receptor  is believed to work in a second stage  as a  coincidence detector,  involved in long-term potentiation and memory.
\end{abstract}

\maketitle

\section{Introduction}
Biological cells live in  very noisy  environments  from which they receive many different stimuli.
The survival of a cell highly depends  on its ability to respond to such stimuli  and to adapt
to changes in the environment. A fundamental role in this task is played by membrane receptors, 
proteins on the cell surface which are able to bind external ligand molecules and trigger  signal transduction pathways.
Although the precision with which  a cell can   measure the concentration of a specific ligand  is  negatively affected by many sources of noise \cite{noise1,noise2,noise3,noise4},
 several examples exist in which
such measurements are performed with surprisingly high  accuracy.
In bacterial chemotaxis for instance, 
the bacterium {\it Escherichia coli} can respond to changes in concentration as low as 3.2 nM \cite{ecoli1}, corresponding to only three molecules in the volume of a cell.
 High accuracy is observed also in spatial sensing by single cell  eukaryotic organisms as  e.g.  in  the slime mold {\it Dictyostelium discoideum},  which is able to sense a concentration difference of only  $1-5\% $ across the cell diameter \cite{dicty1},  and in {\it Saccharomyces cerevisiae} (budding yeast), which is able to orient growth in a gradient of $\alpha$-pheromone mating factor down to estimated $1\%$ receptor occupancy difference across the cell \cite{segall93}.
Spatial sensing is also efficiently performed  by  growing neurons, lymphocytes, neutrophils and other cells of the immune system.
\begin{figure}[hth]
\includegraphics[height=9.4 cm,width=6.8 cm, angle=90]{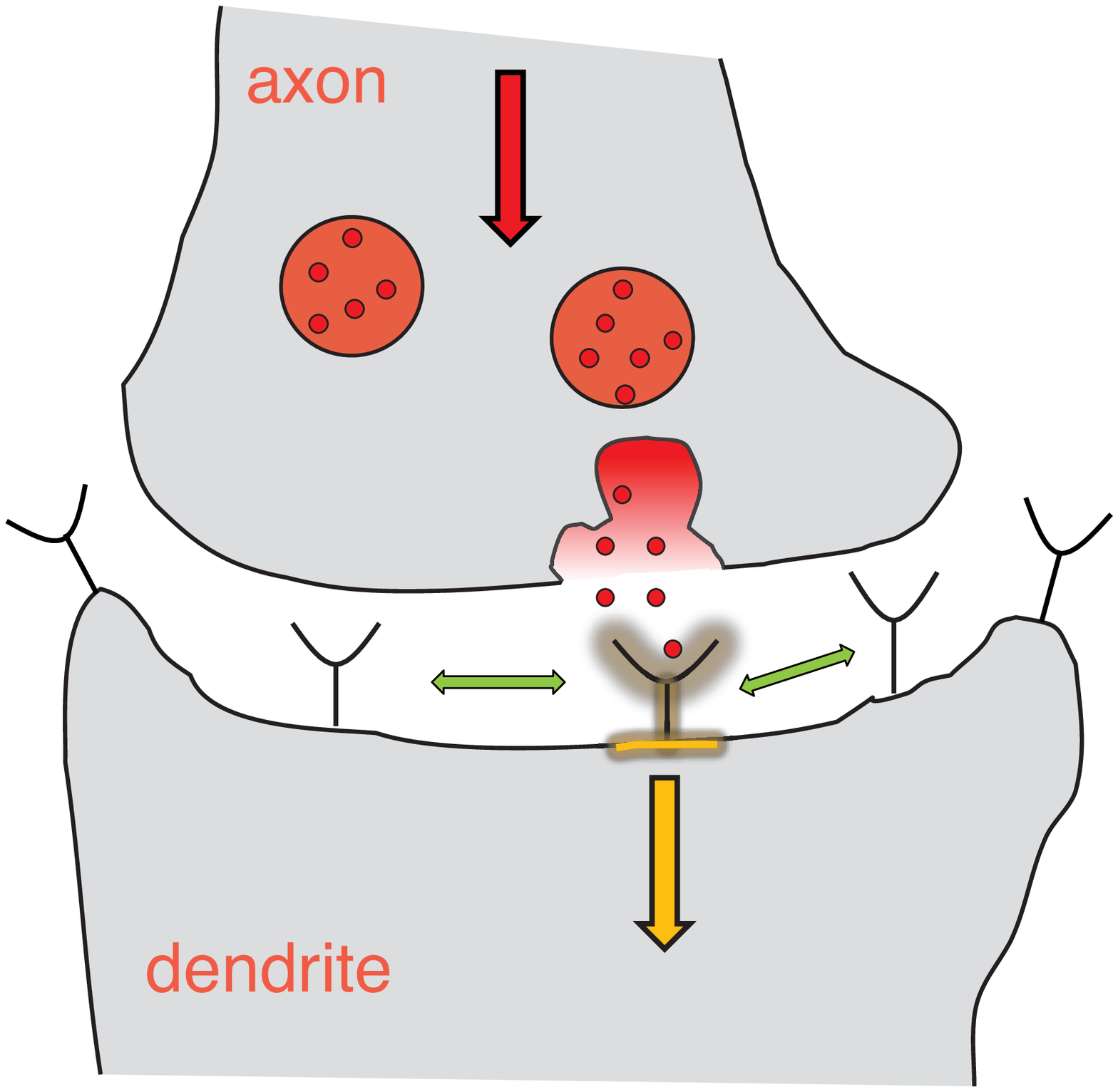}
%
%
\caption{(Color online) Synaptic junction during neurotransmission. A vesicle containing  a neurotransmitter (e.g. glutamate for excitatory synapses), releases its content into the synaptic cleft. Receptors on the post-synaptic surface
bind the neurotransmitter,  increasing the likelihood of 
the propagation of the electric impulse (action potential). Red arrow indicates the incoming potential, yellow arrow the set of  reactions leading to propagation of potential,  small green arrows indicate  receptor mobility.}
\label{fig_00}
\end{figure}

There has been substantial progress  in our theoretical  understanding of the accuracy of concentration sensing. 
  In 1977,
 Berg and Purcell were the first to point out that  ligand sensing is limited by 
ligand 
diffusion,
producing a ``counting noise'' at the receptors \cite{bergP}.  Bialek and Setayeshgar \cite{bialek} (and later others \cite{robned08}) used the Fluctuation Dissipation Theorem (FDT) to separate the contributions from random binding/unbinding events from rebinding due to diffusion of ligand molecules.
Subsequently, Wingreen and Endres 
 showed that the perfect absorber is the  absolute fundamental limit of ligand sensing accuracy; it  does not suffer from  ligand unbinding and rebinding noise \cite{robned08, robnedprl}.

In  previous work   we  analyzed the contribution of endocytosis,  i.e. the internalization of  cell-surface receptors, to the accuracy of external ligand concentration  sensing \cite{int1}.
We showed with a simple model that internalization, by making the cell act as an absorber of ligand, increases sensing accuracy by reducing the noise from  rebinding of already measured ligand molecules.
In this paper we consider the effect of
lateral diffusion of  receptors on the accuracy of  ligand concentration sensing.

 In a situation in which ligand sensing is concentrated  in a specific region on the cell membrane (signaling hotspot),
receptors  bind ligand molecules locally  but may  release them remotely from the region of interest, thus preventing those particles from rebinding locally. Noise from overcounting could  therefore be  reduced, potentially increasing the  
 accuracy of sensing.

A potential  implementation of such localized sensing occurs in neural synapses, responsible for transmission of action potentials and short-term synaptic plasticity (see Fig. 1 for a schematic description).
At such critical places in the central nervous system, the accuracy of sensing ought to be important \cite{aldo}.
At a synapse the propagation of an action potential arriving from the pre-synaptic neuron causes the opening of ion channels on the pre-synaptic membrane which increases calcium-ion inflow \cite{cellbook}. Calcium ions  trigger a biochemical cascade which results in 
the  formation of vesicles containing neurotransmitter, e.g. glutamate, in the case of excitatory synapses.
One or more  vesicles eventually fuse with the membrane   
at a  point on the pre-synaptic surface to  form a fusion pore. Glutamate is subsequently released and  diffuses into the synaptic cleft, reaching the receptors on the post-synaptic surface on the opposite side of the cleft. These receptors, mainly the  AMPA and NMDA receptors (named after the agonists $\alpha$-amino-3-hydroxy-5-methyl-4-isoxazolepropionic acid and N-methyl-D-aspartic acid respectively),  cause ion channels on the post-synaptic membrane to open, 
 facilitating the propagation of  the action potential.

The AMPA receptor (AMPAR) has a tetrameric structure with  four  binding sites for glutamate. Binding glutamate on two sites  leads to conformational change and opening of a pore.
 More occupied binding sites  imply a higher current through the channel, mainly of calcium,  sodium and  potassium  ions. 
The NMDA receptor (NMDAR) has a tetrameric structure as well, but its channel 
needs coincidental binding of glutamate and glycine molecules on   dedicated sites for it to open.
A small membrane depolarization is furthermore necessary to clear out the magnesium ions
blocking the channel. This third condition makes the NMDA receptor a type of coincidence detector  for membrane depolarization and synaptic transmission, playing an  important role in  memory and learning.
Interestingly, AMPA receptors  diffuse on the post-synaptic surface unusually fast, especially when ligand is bound \cite{renner}. In addition to refreshing the synaptic plasticity \cite{heine}, such diffusion may also increase the accuracy of sensing in neurotransmission.

 This paper is organized as follows: in Sec. II we review the case of a single immobile receptor; in Sec. III we consider  receptor diffusion on the membrane and derive the dynamical equations determining the   spatial concentration of ligand and  the occupancy of the receptor.
In Sec. IV we derive  the stationary solution for the receptor occupancy, and in Sec. V, using a non-equilibrium 
approach based on the effective temperature, we derive the accuracy of sensing. In Section VI we describe how our theoretical results connect to 
 neurotransmission in neural synapses.
The final section is devoted to an overall discussion
including cases of further biological relevance.
Technical details are provided in two appendices.


\section{Review of the single receptor}
 In this section we  review previous results for a single, immobile receptor.
 As  depicted in Fig. 2, such a  receptor can  bind and release ligand with rates $k_+ \cb$ and 
$k_-$, respectively. The kinetics for the occupancy $n(t)$ of the receptor are therefore given by
\BEQ
\label{basicsingle}
\frac{d n(t)}{d t}=k_{+} \bar{c}\left[1-n(t)\right] -k_- n(t),
\EEQ
where   the concentration of ligand, $\cb$, is assumed uniform and constant.
The steady-state solution for the receptor occupancy is given by
\BEQ
\label{nbKd}
\nb=\frac{\cb}{\cb+K_D}
\EEQ
with $K_D=k_-/k_+$ the ligand dissociation constant.
The rates of binding and unbinding are related to the (negative) free energy $F$ of binding through the detailed balance
\BEQ
\frac{k_+ \cb}{k_-}=e^{\frac{F}{T}}
\EEQ
with $T$ the temperature in energy units.
In  the limit of very fast ligand diffusion,  i.e.  when
a ligand molecule is immediately removed from the receptor after unbinding,
the receptor dynamics are effectively decoupled from
the diffusion  of ligand molecules and hence, diffusion does not need to be included explicitly.

Following Bialek and Setayeshgar \cite{bialek}, the accuracy of sensing is obtained  by applying the FDT \cite{kubo}, which relates
the spectrum of fluctuations in occupancy to the linear response of the receptor occupancy to a perturbation in the 
receptor binding energy. Furthermore,  at equilibrium 
the fluctuations in occupancy can be directly related to the uncertainty in ligand concentration using Eq. (\ref{nbKd}). After time-averaging over a duration $\tau$  much larger than the correlation time of the binding and unbinding events, the normalized uncertainty of sensing is given by \cite{bialek,rob}
\BEQ
\label{acc1}
\frac{\langle (\delta c)^2\rangle_{\tau}}{\cb^2}=  \frac{2}{k_+ \cb (1-\nb) \tau} \to \frac{1}{2 \pi D_3 \cb s \tau},
\EEQ
where 
 the right-hand side is obtained for diffusion-limited binding \cite{rob},  i.e. when $k_+ \cb (1-\nb) \to 4 \pi \cb  D_3 s$, with $D_3$ the diffusion constant and $s$ the dimension of the (spherical) receptor.
Eq. (\ref{acc1}) shows that the uncertainty is limited by the random binding and unbinding of ligand. The accuracy of sensing is defined as the inverse of the uncertainty.

\begin{figure}[hth]
\includegraphics[width=3.9 cm,angle=0]{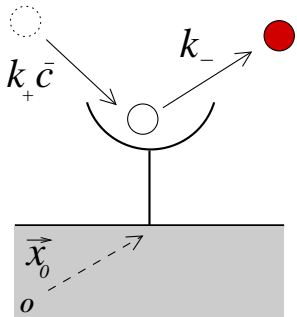}
\caption{(Color online) Single receptor, immobile at position $\vec{x}_0$,  binds and unbinds ligand  with rates $k_+ \cb$ and 
$k_-$, respectively.}
\label{fig_00}
\end{figure}
In  the case where  diffusion of ligand
is slow,   ligand binding to the receptor is 
perturbed 
  \cite{bialek}.
The kinetics of  receptor occupancy and ligand concentration are 
described by
\begin{subequations}
\begin{align}
&\frac{d n(t)}{d t}=k_{+} c(\vec{x}_0,t)[1- n(t)] -k_-n(t) \label{2Dcoupleda}\\
 &\frac{\partial c(\vec{x},t)}{\partial t}=D_3 \nabla^2 c(\vec{x},t)-\delta(\vec{x}-\vec{x}_0)\frac{d n(t)}{d t} \label{2Dcoupledb},
\end{align}
\end{subequations}
where
 $\vec{x}_0$  indicates the position of the receptor. In the last term in the second equation, the Dirac delta function $\delta(x-\vec{x}_0)$
describes a sink or source of  ligand
at $\vec{x}_0$ 
corresponding to ligand binding or unbinding from the receptor, respectively.

Following a similar procedure as in the previous case,
the uncertainty of sensing is given by \cite{bialek,rob}
\begin{subequations}
\label{dc3D}
\begin{align}
\frac{\langle (\delta c)^2\rangle_{\tau}}{\cb^2}
&= \frac{2 }{k_+ \cb (1-\nb) \tau} +\frac{1}{ \pi  D_3 \cb s\tau} \label{dc3Da} \\
&\to \frac{3}{ 2 \pi  D_3 \cb s\tau}, \label{dc3Db}
\end{align}
\end{subequations}
where the first term on the right-hand side of Eq.\ (\ref{dc3Da}) is the same as in Eq.\ (\ref{acc1}),
while the second term is 
the increase in uncertainty due to diffusion.
This  term accounts for the additional  measurement uncertainty from rebinding of  previously bound ligand to the receptor.
For diffusion-limited binding, one obtains  Eq.\ (\ref{dc3Db}) \cite{rob}.

Comparison of Eqs.\ (\ref{acc1}) and (\ref{dc3D}) shows that removal of previously bound ligand by  fast  diffusion
increases the accuracy of sensing, since the same ligand molecule is never measured more than once.

The uncertainty  in Eq. (\ref{acc1}) can be  further reduced by a factor of two, consequently increasing the accuracy of sensing by the same factor,  
  by considering the fundamental limit. In this limit, ligand particles are absorbed, and hence only binding noise contributes. For diffusion-limited binding, the fundamental limit is given by \cite{bergP,robned08,robnedprl}
\BEQ
\label{fundlimit}
\frac{\langle (\delta c)^2\rangle_{\tau}}{\cb^2}= \frac{1}{ 4 \pi D_3 \cb s\tau},
\EEQ
calculated from the diffusive flux to a small sphere of radius $s$, which represents the receptor.

\section{Receptor Diffusion Model}
In  this model, depicted in Fig. 3,  receptors are free to diffuse
on the 2-dimensional (2D) membrane surface with diffusion coefficient $D_2$.  A receptor located in the limited area $\Delta A$ centered at $\vec{x}_0$, can bind and unbind ligand of average concentration $\cb$ 
with rates $k_+ \cb$ and $k_-$, respectively. 
When a receptor  leaves this area and diffuses to the remaining large region of negligible ligand concentration, it can either release ligand or diffuse back in the area.

We consider a situation in which  receptor diffusion  has  reached a stationary condition resulting  in a uniform density $\rho_0$ of receptors on the membrane, and this condition is not changed by  interaction with ligand molecules.
Ligand binding and unbinding only affects the number of occupied and unoccupied receptors with respective densities $\rho_b$ and $\rho_u$, given by
\begin{figure}[t]
\includegraphics[width=6.4 cm,angle=0 ]{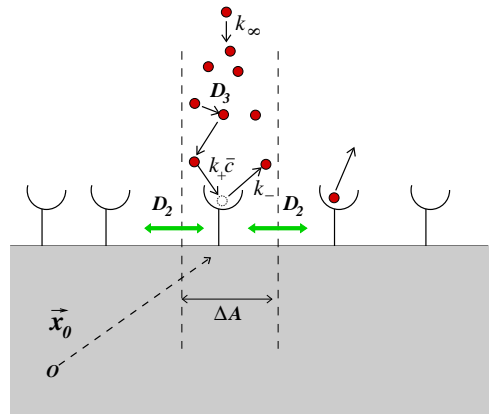}
\caption{(Color online) Ligand-receptor binding by mobile receptors characterized by diffusion coefficient $D_2$ (green arrows).
 Ligand (red dots) is concentrated only 
in a small area $\Delta A$ around $\vec{x}_0$ due to  e.g.  local release of glutamate from
vesicle fusion pore with rate $k_{\infty}$. Ligand can bind to receptors at $\vec{x}_0$, but unbind everywhere on the membrane. Parameter $D_3$ describes the diffusion coefficient of the ligand.}
\label{fig_00}
\vspace{0.2cm}
\end{figure}
\BEQ
\label{ro0}
\rho_0=\rho_b(\vec{r}) +\rho_u(\vec{r}),
\EEQ
where $\vec{r}=(x,y)$  is a position  on the 2D membrane  surface.
 For purposes of comparison with  the single immobile receptor model, we 
 also assume  that there is always  exactly  one receptor in the area $\Delta A$  around $\vec{x}_0=(\vec{r}_0, z_0)$, i.e. $\rho_0=1/\Delta$.

The dynamics of such a system is captured by  the following set of coupled differential equations
\begin{subequations}
\label{dyn1}
\begin{flalign}
\label{dyn1a}
\frac{\partial \rho_b(\vec{r},t)}{\partial t}&=D_2 \nabla^2 \rho_b(\vec{r},t)-k_-\rho_b(\vec{r},t)+ \\
\nn &k_{+}[\rho_0- \rho_b(\vec{r},t)]  \,  c(\vec{r},z_0,t)\, \Delta A \, \delta(\vec{r}-\vec{r_0})\\
\nn \frac{\partial c(\vec{x},t)}{\partial t}&=D_3 \nabla^2 c(\vec{x},t)-\delta(z-z_0) \left[ \frac{\partial \rho_b(\vec{r},t)}{\partial t}  \right.\\
& \left. -D_2 \nabla^2 \rho_b(\vec{r},t)\rule{0 cm} {0.5 cm}\right] +k_{\infty} \delta(\vec{x}-\vec{x}_{\infty}), \label{dyn1b}
\end{flalign}
\end{subequations}
where  rate constant $k_{\infty}$ ensures the replenishment of ligand molecules,
$\vec{x}=(x,y,z)=(\vec{r},z)$ indicates a  location  in 3D space. Specifically,
 $\rho_b(\vec{r},t)$ is the density of bound receptors and 
$c(\vec{x},t)$ is the ligand concentration, both depending on space and time.
An additional equation for $\rho_u(\vec{r},t)$ is  not necessary due to receptor conservation in Eq. (8).

After normalization by the average density, using $n(\vec{r},t)=\rho_b(\vec{r},t)/\rho_0$ in  Eq. (\ref{dyn1}), we obtain
\begin{widetext}
\begin{subequations}
\label{dyn2}
\begin{align}
\label{dyn2a}
&\frac{\partial n(\vec{r},t)}{\partial t}=D_2 \nabla^2 n(\vec{r},t)+k_{+}[1- n(\vec{r},t)]   \,  c(\vec{r},z=z_0,t)\, \Delta A \,\delta(\vec{r}-\vec{r_0})-k_-n(\vec{r},t)\\
&\frac{\partial c(\vec{x},t)}{\partial t}=D_3 \nabla^2 c(\vec{x},t)-\rho_0 \delta(z-z_0)\left[\frac{\partial n(\vec{r},t)}{\partial t}-D_2 \nabla^2 n(\vec{r},t)\right]+k_{\infty} \delta(\vec{x}-\vec{x}_{\infty}).\label{dyn2b}
\end{align}
\end{subequations}
\end{widetext}

These equations describe the coupled dynamics of the diffusing receptors and their  
 occupancy with ligand. In summary, receptors can bind ligand molecules only within area  $\Delta A$ 
but can release them anywhere on the membrane.
The solution is provided in the following sections.

\section{Non-equilibrium Stationary solution for receptor occupancy}
It is  possible to extract an exact analytical expression for the stationary solution
$\nb(\vec{r})$  for the receptor occupancy from  Eq. (\ref{dyn2a}).
Setting the right-hand side to zero and Fourier Transforming both sides leads to
\begin{align}
\hat{\bar{n}}(\vec{q})=\frac{k_+ [1-\nb(\vec{r}_0)] \cb(\vec{x}_0) \Delta A}{D_2 q^2 +k_-},
\end{align}
where $\vec{q}$ is the wave  vector of magnitude $q=|\vec{q}|$.
The inverse Fourier Transform, upon introducing a cut-off $\Lambda\sim 1/s$ due to the receptor size, leads to the following solution
\begin{align}
\label{exact}
\nb(\vec{r})= k_+ [1-\nb(\vec{r}_0)] \cb(\vec{x}_0) \frac{\Delta A}{4 \pi D_2}\log\left(1+\frac{D_2 \Lambda^2}{k_-}\right).
\end{align}
 This stationary solution corresponds to a non-equilibrium condition since
ligand is continuously taken away from area $\Delta A$.
Setting $\Delta A=4 \pi/\Lambda^2 \sim s^2$ and evaluating Eq. (\ref{exact})  in $\vec{r}_0$ leads therefore to the following value for the occupancy at $\vec{r}_0$ 
%
%
\begin{align}
\label{bess2}
\nb(\vec{r}_0)=\frac{1
}{1+\left[ k_+  \cb(\vec{x}_0)  \frac{\Delta A}{4 \pi D_2}\log\left(1+\frac{4 \pi D_2}{\Delta A k_-}\right)\right]^{-1}},
\end{align}
which  is plotted in Fig. 4 and, in the limit $D_2\to 0$, leads to the recovery of the immobile receptor
solution  Eq. (\ref{nbKd}), namely  $\nb_0=k_+ \cb/(k_-+k_+ \cb)$.
\begin{figure}[hth]
\includegraphics[width=8.8 cm, angle=0]{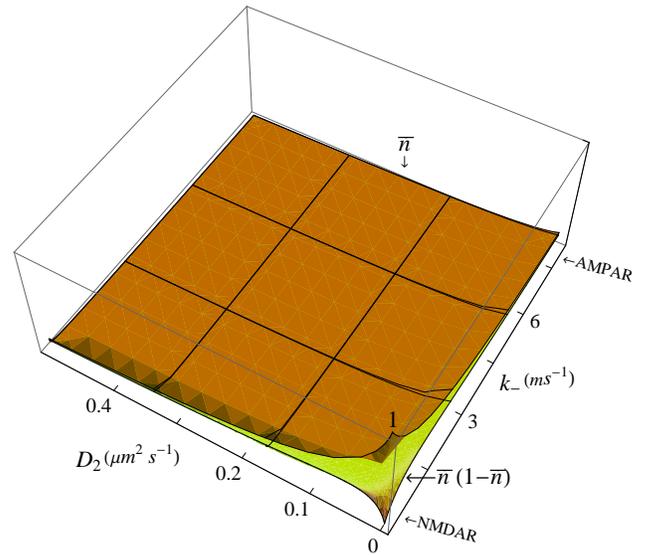}
\caption{(Color online) Receptor occupancy at $\vec{r}_0$.
 Eq. (\ref{bess2}) for $\nb$ is plotted as a function of lateral diffusion coefficient $D_2$ and  unbinding rate $k_-$ for a range of experimentally relevant values in neurotransmission ({\it top surface}). Specifically, the  $k_-$ values for AMPA and NMDA receptors are indicated  (see Table I).  Corresponding plot for $\nb(1-\nb)$ is also shown ({\it bottom surface}). }
\end{figure}
In order to quantify the deviation from equilibrium 
we apply  Fick's first law to calculate the flux $\vec{J}$ of 
occupied receptors 
\begin{align}
\label{fick1}
\vec{J}=D_2 \vec{\nabla} \nb(\vec{r},t)
\end{align}
out of region $\Delta A$ \cite{bergbook}.
Integrating  Eq. (\ref{fick1}) over the border of $\Delta A$ 
allows us to define an effective
``internalization rate'' for removal of ligand  in analogy to \cite{int1}, via  
\begin{align}
\nn \kf \nb&= -\frac{D_2}{\Delta A} \oint \nabla \nb (\vec{r}) d \vec{r}=-\frac{D_2}{\Delta A}\int_{\Delta A} \nabla^2 \nb(\vec{r}) d^2 \vec{r}\\
&= \frac{1}{\Delta A} \int_{\Delta A} d^2 \vec{r} \left[k_+ \cb (1-\nb) \Delta A \delta (\vec{r}) -k_- \nb \right],
\end{align} 
where the  second equality follows from integration by parts and the last equality results
from the stationary condition for Eq. (\ref{dyn2a}).
Carrying out the last integration leads to the following final expression for the effective internalization rate
\begin{align}
\label{keff}
\kf = k_+ \cb \left( \frac{1-\nb}{\nb}\right) -k_-=\frac{4 \pi D_2}{\Delta A\log\left(1+\frac{4 \pi D_2}{\Delta A k_-}\right)}-k_-,
\end{align}
where we used Eq. (\ref{bess2}) to express $\nb$ in terms of the receptor diffusion coefficient.
Using condition  $\Delta A=4 \pi/\Lambda^2$ we can define a  modified unbinding rate
\BEQ
\kappa_-=k_-+\kf=\frac{k_+ \cb (1-\nb)}{\nb}=\frac{ D_2 \Lambda^2}{\log\left(1+\frac{D_2 \Lambda^2}{ k_-}\right)}
\EEQ
and  extend the equilibrium condition Eq. (3) to the stationary non-equilibrium condition  via an effective temperature $T_e$ defined by \cite{int1}
\BEQ
\label{eff1}
\frac{k_+ \cb}{\kappa_-}=e^{\frac{F}{T_e}}.
\EEQ
This effective temperature allows us to generalize  the ordinary FDT to a stationary non-equilibrium condition and  to derive
 the receptor accuracy of sensing in the next section.

\section{Accuracy of Sensing}

Here we
we derive  the accuracy of sensing for diffusing receptors.
Starting   from Eq. (\ref{dyn2}) we consider an 
expansion to first order around the stationary solutions
\begin{subequations}
\begin{align}
 n(\vec{r},t)&=\bar{n}+\delta n(\vec{r},t) \label{nca}\\
c(\vec{x},t)&=\bar{c}+\delta c(\vec{x},t),\label{ncb}
\end{align}
\end{subequations}
where $\nb$ and $\cb$ are the stationary solutions of Eq. (\ref{dyn2}), which, in this case,
have a spatial dependence,  i.e. $\nb=\nb(\vec{r})$ and $\cb=\cb(\vec{x})$.  Hence, the linearized equation for the receptor occupancy and ligand concentration are given by 
\begin{widetext}
\begin{subequations}
\label{linear1}
\begin{align}
\label{linear1a}
\frac{ \partial [\dnr]}{\partial t}&=D_2 \nabla^2 \dnr+ \left[ (1- \bar{n})   \delta c(\vec{r},z_0,t) - \bar{c} \dnr   +(1- \bar{n}) \bar{c} \frac{\delta k_+(t)}{k_+} -\bar{n} \frac{\delta k_-(t)}{k_+} \right]k_+ \Delta A \, \delta(\vec{r}-\vec{r_0})-k_- \dnr \\
\label{linear1b}\frac{\partial [\dc] }{\partial t}&=D_3 \nabla^2 \dc -\delta(z-z_0)\rho_0 \Bigg\{\frac{\partial \left[ \dnr \right]}{\partial t}-D_2 \nabla^2 \dnr \Bigg\},
\end{align}
\end{subequations}
\end{widetext}
where we  assumed that fluctuations in the binding/unbinding rate constants occur as well. This mathematical trick allows us to introduce fluctuations in the receptor-binding free energy and so to apply the FDT  \cite{bialek, kubo}.
Using  the stationary non-equilibrium condition  Eq. (\ref{eff1}), we obtain the fluctuations in the rates and binding free energy
%
\BEQ
\label{ft1}
\frac{\delta k_+}{k_+}- \frac{\delta k_- }{\kappa_-}=\frac{\delta F}{T_e}.
\EEQ  

Using this equation allows us to replace  fluctuations in the rate constants  in  Eq. (\ref{linear1a}) with fluctuations in the binding free energy, resulting in 
\begin{flalign}
\label{linearf}
\nn &\frac{ \partial [\dnr]}{\partial t}=D_2 \nabla^2 \dnr -k_- \dnr +\left[  k_+ \cb (1-\nb) \frac{\delta F}{T_e} \right.\\
 & \left. + k_{+}(1- \bar{n}) \, \delta c(\vec{r},z_0,t) 
-k_+ \bar{c} \delta n(\vec{r},t) \rule{0 cm} {0.5 cm}\right] \Delta A \, \delta(\vec{r}-\vec{r_0}).
\end{flalign}

 Fourier Transforming Eq. ({\ref{linearf}) and setting
  $q=|\vec{q}|$, we solve for
 $\dnhq$ and obtain
\BEQ
\label{nqw}
\dnhq =G(\omega,\vec{r}_0,z_0) \frac{e^{i\vec{q}\cdot \vec{r_0}}}{D_2 q^2 +k_--i \omega },
\EEQ
where, for convenience of calculation, we have defined the following function
\begin{flalign}
\label{Gdef}
\nn G(\omega,\vec{r}_0,z_0)&=\left[\rule{0 cm} {0.5 cm} k_+ \rnbo \, \dchz -k_+ \cb \,\dnhz+ \right. \\
 & \left. \rnbo \cb \frac{\delta F(\omega)}{T_e}\right]\Delta A.
\end{flalign}
Inverse Fourier Transforming  Eq. (\ref{nqw}) leads to the following expression for the spectrum of the
fluctuations in occupancy
\BEA
\label{nzw}
\dnhz&=& \int \frac{d \vec{q}}{(2 \pi)^2}e^{-i \vec{q} \cdot \vec{r}_0}\dnhq\\
\nn &=&G(\omega,\vec{r}_0,z_0)
 \int \frac{d \vec{q}}{(2 \pi)^2} \frac{1}{D_2 q^2 +k_--i \omega }\\
\nn&=&G(\omega,\vec{r}_0,z_0)\Sigma_1(\omega),
\EEA
where we defined
\BEQ
\label{sigma1Int}
\Sigma_1(\omega)=\int \frac{d \vec{q}}{(2 \pi)^2} \frac{1}{D_2 q^2 +k_--i \omega }
\EEQ
with the integral provided in Appendix A.
In order to remove the ligand concentration in Eq. (\ref{Gdef}), we 
Fourier Transform Eq. (\ref{linear1b}), leading to
\BEQ\label{ceq}
\dch=\rho_0\frac{ i \omega -D_2 q^2}{D_3 (q^2+ q_{\perp}^2)-i \omega } \dnhq e^{i q_{\perp} z_0}.
\EEQ
Inserting Eq.  (\ref{nqw}) in Eq. (\ref{ceq}) and inverse Fourier Transforming leads to the following expression for the spectrum of the
fluctuations in  ligand concentration
\BEA
\label{czw}
\nn \dchz&=&\int \frac{dq_{\perp}}{2\pi}e^{-i q_{\perp} z_0} \int \frac{d \vec{q}}{(2 \pi)^2}e^{-i \vec{q} \cdot \vec{r}_0}\dch\\
&=&G(\omega,\vec{r}_0,z_0)\rho_0\Sigma_2(\omega)
\EEA
with 
\BEQ
\label{sigma2Int}
\Sigma_2(\omega)=\int \frac{dq_{\perp}}{2\pi}\int \frac{d \vec{q}}{(2 \pi)^2}\frac{i \omega/D_2 - q^2}{D_3(q^2+q_{\perp}^2)-i \omega} \frac{e^{-i \vec{q} \cdot \vec{r}_0}}{q^2 +\frac{k_-}{D_2} -\frac{i\omega}{D_2}}
\EEQ
provided in Appendix B.
Inserting Eqs. (\ref{nzw}) and (\ref{czw}) 
  in Eq.  (\ref{Gdef}) leads to  a closed equation
for the function $G(\omega,\vec{r}_0,z_0)$
\begin{flalign}
&G(\omega,\vec{r}_0,z_0)=\left[\rule{0 cm} {0.5 cm} k_+ \rnbo  \rho_0 \Sigma_2(\omega)G(\omega,\vec{r}_0,z_0) + \right.\\
\nn  & \left. -k_- \cb \Sigma_1(\omega)  G(\omega,\vec{r}_0,z_0) +k_+ \cb \rnbo \frac{\delta \hat{F}(\omega)}{T_e}\right] \Delta A,
\end{flalign}
which can be solved for $G(\omega,\vec{r}_0,z_0)$. This leads to
\BEQ
\label{Gexpl}
G(\omega,\vec{r}_0,z_0)=  \frac{k_+ \cb \,  \rnbo \Delta A \,\, \,\delta \hat{F}(\omega)/T_e }{1-k_+ \rnbo \Sigma_2(\omega) +k_+ \cb \Sigma_1(\omega) \Delta A},
\EEQ
where  condition $\rho_0=1/\Delta A$ was applied.
Using
 Eqs. (\ref{nzw}) and (\ref{Gexpl}), a final expression
for the response is obtained
\BEA
\label{response}
\frac{\dnhz}{\delta \hat{F}(\omega)}&=&\frac{G(\omega,\vec{r}_0,z_0) \Sigma_1(\omega)}{\delta \hat{F}(\omega)}\\
 \nn &=&\frac{1}{T_e} \frac{k_+ \cb \rnbo  \Delta A \,\Sigma_1(\omega)}{1-k_+ \rnbo \Sigma_2(\omega) +k_+ \cb \Sigma_1(\omega) \Delta A}.
\EEA

\begin{center}
\begin{figure}
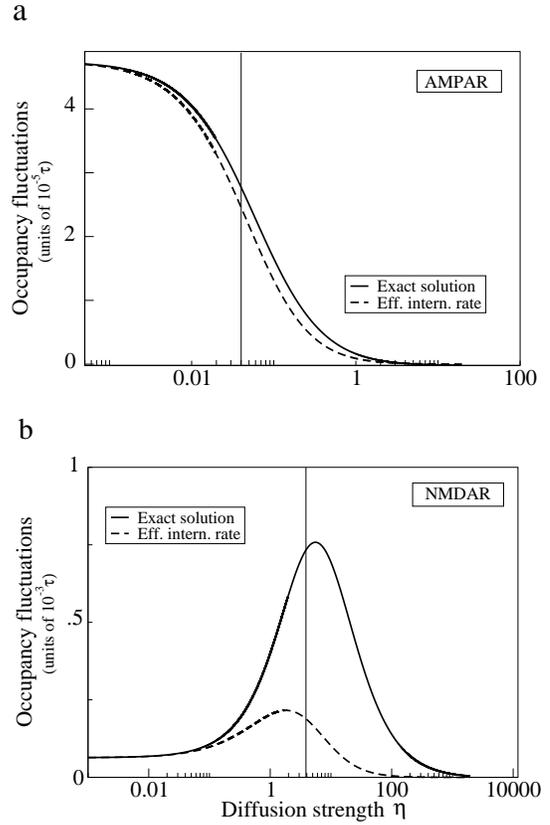

    \begin{tabular}{ccc}
&     \resizebox{70mm}{!}{\includegraphics{fig5a.eps}} 
\vspace{0.3cm}
\\
 &     \resizebox{70mm}{!}{\includegraphics{fig5b.eps}}
 \end{tabular}
\caption{Receptor-occupancy  fluctuations  $\langle (\delta n)^2\rangle_{\tau}$   in units of $\tau$  as  a function of $\eta=16 D_2/(s^2 k_-)$ for AMPAR (a) and NMDAR (b). Shown are ({\it solid line}) plot of $\langle (\delta n)^2\rangle_{\tau}$ as given by Eqs. (\ref{exactF}) and (\ref{deltan}),
({\it  dashed line})  plot of Eq. (\ref{deltanint})  using the effective internalization rate Eq. (\ref{keff}),  and
({\it vertical thin line})  experimental  value of $\eta$
for each receptor type (see Table I for all parameters value). Glutamate concentration  was set to $\cb$=0.1mM, here and in Fig. 6. In order to produce a significant effect from ligand rebinding, we used $D_3=0.1 \mu m^2/s$ here and in subsequent plots, as may result from binding of glutamate to neuroligins, neurexins and other cleft proteins.}
 \label{test4}
 \end{figure}
 \end{center}

Applying the generalized  FDT
we derive the spectrum of the fluctuations in receptor occupancy $\dnhz$ from the response, namely
\BEQ
\label{gFDT}
S(\omega)=\frac{2 k T_e}{\omega} \text{Im}\left[\frac{\dnhz}{\delta \hat{F}(\omega)}\right].
\EEQ
By replacing the expressions for $\Sigma_1(\omega)$ and $\Sigma_2(\omega)$  in Eq. (\ref{response}), taking the imaginary part and setting $\omega \simeq 0$,
 we obtain for the zero-frequency limit of the power spectrum
\begin{flalign}
\label{exactF}
\nn&
S_{\eta}(0)=\frac{k_+ \cb \rnbo \Delta A \Lambda^2}{2 \pi k_-^2 (1+\eta)
\left[1+ \frac{k_+ \cb}{k_-}L_{\eta}+\frac{k_+ (1-\nb)\Lambda}{8 \pi D_3}\left(1-A_{\eta}\right)\right]^2} \times \\
&
 \left[1+\frac{k_+ (1-\nb)\Lambda}{16 \pi D_3}\left(1+\frac{D_3 \Lambda^2-\eta k_-}{D_3 \Lambda^2/(1+\eta)}A_{\eta}+\frac{1-A_{\eta}}{L_{\eta}/2}\right) L_{\eta}
 \right],
\end{flalign}
where we  conveniently defined the following functions of the dimensionless
parameter $\eta=D_2 \Lambda^2/k_-$
\begin{flalign}
 L_{\eta}\equiv \frac{\log\left(1+\eta \right)}{\eta}\;\;,\;\;\; A_{\eta}\equiv\frac{\arctan{\sqrt{\eta}}}{\sqrt{\eta}}.
\end{flalign}
Note that in Eq. (\ref{exactF}), occupancy $\nb$ also depends on $\eta$ via Eq. (\ref{bess2}).
The zero-frequency spectrum  is related to the receptor occupancy fluctuations, averaged over a time $\tau \gg k_-^{-1},( k_+ \cb)^{-1}$, by the following relation:
\BEQ
\label{deltan}
\langle (\delta n)^2 \rangle_{\tau}\simeq S_{\eta}(0)/\tau.
\EEQ
 In Fig. 5 we plot  the occupancy fluctuations, given by Eq. (\ref{deltan}) as a function of $\eta$, i.e.  the effective  diffusion strength. This is done for the two main receptors involved in neurotransmission, namely AMPAR and  NMDAR, with parameters taken from Table I.  
For AMPAR, the effect of diffusion is that of decreasing  the fluctuations in occupancy
due to reduced overcounting of previously bound molecules
(Fig. 5a).
For small unbinding rates $k_-$ though, such as occurs for the  NMDA receptor, an actual increase in the noise is observed
for a range of  physiologically relevant diffusion 
strengths.
\begin{center}
\begin{table}[ht!]
\begin{tabular}{|l||l|l|}
\hline
 \; \; & \multicolumn{2}{c|}{ GLUTAMATE} \\
\hline
$ D_3$ & \multicolumn{2}{c|}{300 $\mu$m$^2$/s(acqueous solution)\cite{postle} } \\
&         \multicolumn{2}{c|}{ 10$\mu$m$^2$/s (synaptic cleft)\cite{cerebellum,trillerD} }\\
 $\cb$ &\multicolumn{2}{c|}{ 0.1-1 mM \cite{dimaio,postle,jonas0}  }                            \\
$\tau$ & \multicolumn{2}{c|}{ 1 ms  \cite{neurobook}}                          \\
 s &  \multicolumn{2}{c|}{ 8 nm \cite{3Dimg,cellbook} }                         \\
\hline
  & AMPAR &NMDAR\\
\hline
$D_2$  &  0.028-0.1$\mu$m$^2$/s \cite{nmdiff,ampdiff1} &0.021$\mu$m$^2$/s \cite{nmdiff,ampdiff1}  \\
$k_+$  &     4-10$\;\times 10^{6}$M$^{-1}$s$^{-1}$\cite{postle,gibb}      & 5-8.4$\;\times 10^{6}$M$^{-1}$s$^{-1}$\cite{nmda,gibb}  \\
$k_-$  &   2-8$\; \times 10^{3}s^{-1}$  \cite{gibb,postle}          & 5-80 $s^{-1}$ \cite{gibb,nmda}\\
\hline
\end{tabular}
\label{table1}
\caption{Summary of experimentally determined parameters for glutamate ligand (top), AMPA receptor (bottom left) and NMDA receptor (bottom right). }
\end{table}
\end{center}
In order to make the origin of this effect more transparent,  we also plot the occupancy fluctuations  in Fig. 5 as calculated for the case of receptor internalization  \cite{int1}
\BEQ
\label{deltanint}
\langle (\delta n)^2 \rangle_{\tau}\simeq
\frac{2 \nb^2(1-\nb)}{k_+ \cb \tau},
\EEQ
where $\nb= k_+ \cb/(k_+ \cb + k_-+k_i^e)$, with $k_i^e$
 defined in Eq. (\ref{keff}).
 The experimental unbinding rate  of NMDAR  leads to values for $\nb$ 
such that starting from $D_2=0$, the function  $\nb(1-\nb)$ increases, reaches a maximum and then decreases again. Such a maximum is avoided by  AMPAR whose spectrum  monotonically decreases when $D_2$ increases (see Fig. 4).
 The comparison between   receptor diffusion and receptor internalization
in Fig. 5
shows qualitatively similar results, confirming the effective ligand removal by receptor diffusion.


\subsection{Equilibrium limit for vanishing $D_2$}
Here we show  that  Eq. (\ref{exactF})
leads to  the expression for the single receptor obtained in Refs. \cite{bialek,rob}  in the limit for $D_2 \to 0$, i.e. $\eta \to 0$.
Taking the limit and using  $\Delta A=4 \pi/\Lambda^2$, we  indeed obtain
\begin{align}
\label{limitD2}
 \lim_{\eta \to 0}  S_{\eta}(0)&=\frac{2 k_+ \cb (1-\nb_0)}{(k_-+k_+\cb)^2} \left[ 1+\frac{k_+ (1-\nb_0) \, \Lambda}{8 \pi  D_3 }\right],
\end{align}
which is identical to the result for the single receptor for $\Lambda=4/s$.
\begin{center}
\begin{figure}
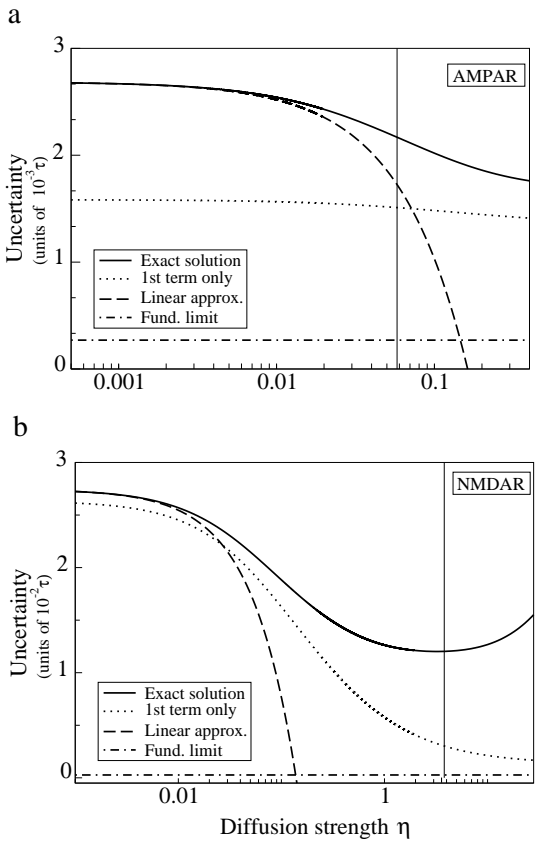

    \begin{tabular}{cc}
&     \resizebox{70mm}{!}{\includegraphics{fig6a.eps}}
 \vspace{0.2cm}\\
 &     \resizebox{70mm}{!}{\includegraphics{fig6b.eps}}
 \end{tabular}
\caption{Uncertainty $\langle(\delta c)^2\rangle_{\tau}/\cb^2$   as a function of $\eta=16 D_2/(s^2 k_-)$ for the AMPA receptor (a) and the NMDA receptor (b) in units of $\tau$.  Shown are ({\it solid line}) plot of Eq. (\ref{accsens0}) as obtained from the exact solution  Eq. (\ref{exactF}), ({\it  dashed line})   linear approximation for small $\eta$, and
 ({\it dotted line}) contribution to the accuracy of sensing from only the first term in Eq. (\ref{accsens}).
Also shown are ({\it vertical thin line})  the experimental value of $\eta$
for each receptor (see Table I) and ({\it horizontal dot-dashed line})  the fundamental physical limit  Eq. (\ref{fundlimit})}
    \label{test4}
 \end{figure}
 \end{center}
\subsection{Near-equilibrium result for small $D_2$}
We derive here the  exact result,  Eq.\ (\ref{exactF}), in the limit of slow receptor diffusion on the membrane, using
$\eta= D_2/(s^2 k_-)$ as a small parameter. 
Specifically, we are interested in the  accuracy of sensing, which  is derived from 
 the  normalized variance for the ligand concentration of ligand as in \cite{bialek}
\BEQ
\label{accsens0}
\frac{\langle (\delta c)^2\rangle_{\tau} }{\cb^2}=\frac{\langle (\delta n)^2 \rangle_{\tau}}{\nb^2(1-\nb)^2 }\simeq \frac{1}{\nb^2(1-\nb)^2 } \frac{S_{\eta}(0)}{\tau},
\EEQ
where the right-hand term  follows from Eq. (\ref{deltan}).
Expansion to first order  in  parameter $\eta$ leads to
\begin{widetext}
\begin{align}
\label{accsens}
 \frac{\langle (\delta c)^2\rangle_{\tau} }{\cb^2}= \frac{2 }{k_+ \cb (1-\nb_0) \tau}\left(1-8 \nb_0\eta\right) +  \frac{1}{ \pi  s D_3 \cb \tau}\left[1-
  \left(\frac{16 k_-^2\nb_0^2}{3\pi k_+\cb^2 D_3 s}+ 8 \nb_0\frac{3k_- -k_+ \cb}{3 k_+ \cb}  +\frac{k_-}{2D_3}s^2 \right) \eta \right].
\end{align}
\end{widetext}
%
Comparison with  Eq.~(6a) 
shows that the correction due to lateral receptor diffusion is twofold to first order: (i)~the first term  is reduced due to a reduction in $\nb$, (ii) the second term is reduced  for most parameter values  due to a reduction in rebinding of already measured ligand molecules.

In Fig.~6 we plot the uncertainty in sensing for both AMPAR and NMDAR.
In both cases the total effect of lateral diffusion is that of decreasing the uncertainty in the concentration and therefore of increasing the accuracy of sensing.
 Receptor diffusion for AMPAR (Fig. 6a)  mainly reduces the rebinding term, while for
   NMDAR (Fig. 6b)    the first term from binding and unbinding  
 is reduced.
In the  latter case,
the maximum in the  occupancy fluctuations (see Fig. 5) is removed  through  $\nb^2 (1-\nb)^2$ in  the denominator of Eq.\ (\ref{accsens0}).
For large values of $\eta$, the occupancy fluctuations and hence the accuracy of sensing become unphysical (zero, i.e. below the physical limit, or diverge) as the effective temperature breaks down far from equilibrium. (For large $\eta$, receptors remove ligand molecules efficiently, which introduces large non-equilibrium ligand fluxes. These cannot be represented by an effective temperature.) 

In summary, lateral  receptor diffusion affects   the accuracy of sensing  in two ways: (i)   by reducing the stationary value for the receptor occupancy, 
 and (ii) by reducing  the noise from rebinding of already measured ligand molecules.
\section{Quantal transmission}
\begin{center}
\begin{figure}
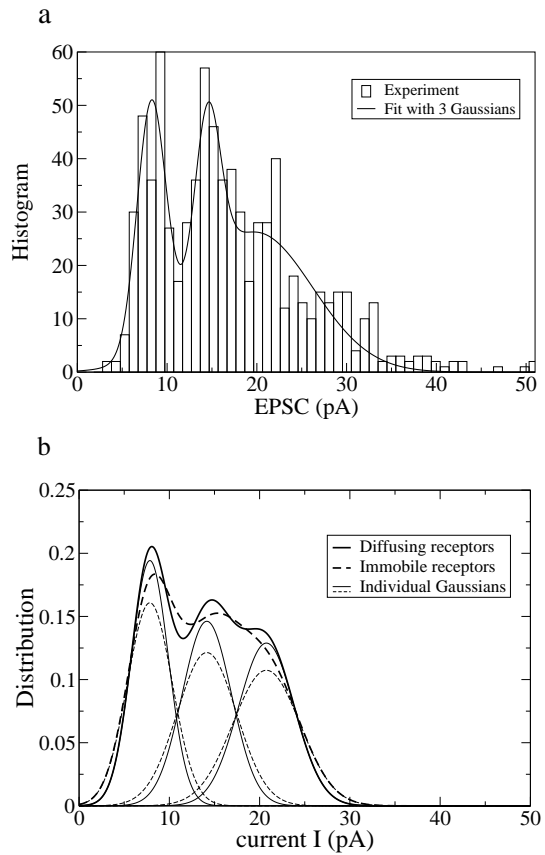

    \begin{tabular}{c}
 \resizebox{70mm}{!}{\includegraphics{fig7a.eps}}
\vspace{0.1 cm}\\
 \resizebox{70mm}{!}{\includegraphics{fig7b.eps}}
 \end{tabular}
\caption{Quantal transmission.  (a)  Histogram of experimental EPSC (excitatory post-synaptic current) data taken from  Ref. \cite{jonas} and  fitted by  three Gaussian distributions. (b) Distribution of predicted EPSCs. Shown are individual Gaussian distributions corresponding to one, two and three vesicles (from left to right; thin lines), as well as envelope functions (sums with equal weighting; thick lines).  Calculation  is done for AMPA receptors using Eqs. (\ref{current}, \ref{deltaI-C}),
Eq. (\ref{accsens0}) for the variances of  diffusing (thin solid lines) and Eq. (6a) for  immobile (thin dashed lines) receptors, assuming glutamate concentration $\cb_n=c_b+n c_v$ for $n=1,2,3$ vesicles, background $c_b=0.02mM$, vesicle concentration $c_v=0.04 mM$ and $N=100$ receptors \cite{kull, cott}.  Furthermore, Eq. (\ref{current}) with parameters from Refs. \cite{jonas,jonas1} was  used.
}
    \label{isto}
 \end{figure}
 \end{center}
A remarkable property of synaptic transmission is that the amplitude of the synaptic response, i. e. the excitatory post-synaptic current (EPSC), varies by an integral multiple
of a quantum. This property, first unveiled by the pioneering work of Katz \cite{katz}, was traced to the formation
of synaptic vesicles and subsequent pore formation. The latter leads to a release of glutamate into the synaptic cleft \cite{heuser}. This quantization is considered a way to minimize the variability of the response from each contributing synapse and therefore to allow efficient neural computation.
This variability is quantified by the coefficient of variation (CV) defined as  the standard deviation devided by the mean value.  The reason for a large CV is controversial (cf. Fig. 7a), but can be attributed to a variation in the size of the vesicles carrying glutamate, or to the fact that large EPSCs  are generated by multiple vesicles \cite{franks, aldo}.  The experimental detection of  EPSCs with an amplitude distribution of  more than one peak  is consistent with the latter explanation \cite{wall}.

In Fig. 7a,  the experimental data taken from Ref. \cite{jonas} show EPSC peaks as  a result of repetitive suprathreshold stimulation of granule cell somata in CA3 pyramidal cells.
Some stimuli did not evoke an EPSC (termed failures).
The histogram of successful events can be fitted with three Gaussian distributions, showing the quantal nature of the EPSC. The peaks of the Gaussians  correspond
 to the discrete multiple values  $7, 14$, and $21$ pA, respectively.

We can compare this  distribution to the result predicted by
the accuracy of sensing. Specifically, we would like to  know if fast diffusion helps resolve the EPSC peaks.
A simple model based on the structural properties of AMPAR
was previously used to derive the dynamics of glutamate binding and unbinding, as well as channel 
opening and closing \cite{jonas}.
Based on this  model 
 the following relation connecting the  current through the ion channel of AMPAR and the concentration of glutamate was deduced:
\BEQ
\label{current}
I=I_0 \frac{1}{1+(\lambda/c)^n},
\EEQ
where $I_0$,  $\lambda$, and the   Hill coefficient $n$ are  parameters provided in Ref. \cite{jonas,jonas1}.
Based on Eq. (\ref{current}) the following relation connecting the width in the distribution of EPSC currents to the fluctuations in glutamate concentration can be derived:
\BEQ
\label{deltaI-C}
\delta I =I_0 n \lambda  \frac{(\lambda/c)^{n-1}}{\left[1+(\lambda/c)^n\right]^2}\delta c.
\EEQ
 We further assume that a number $N$ of receptors independently contribute to the measurement, i. e.
\BEQ
\label{deltacN}
\frac{\langle (\delta c)^2\rangle_{\tau,N}}{\cb^2}=\frac{\langle (\delta c)^2\rangle_{\tau}}{N \cb^2},
\EEQ
with the uncertainty of a single receptor given by Eq. (\ref{accsens0}). Averaging over $N$, therefore, further reduces the uncertainty in Eq.\ (\ref{deltacN}) and hence increases the accuracy of sensing.
 
 Figure 7b shows  that  the distributions of predicted  EPSCs is in qualitative agreement with the experiments. 
 From our calculations we found that the effect of  lateral receptor diffusion  is to increase the resolution of peaks and so also the ability of the synapse to count the number of vesicles released into the synaptic cleft. Note that in order to resolve  the EPSC peaks  experimentally,  the probability of  vesicle release from the presynaptic side was drastically reduced. This was achieved by
modulating  the ratio of Ca$^+$/Mg$^+$ ions in the extracellular solution, 
resulting  in a 50-60\% reduction of peak currents \cite{jonas}.



\section{Discussion and Conclusions}

Our previous work  on the fundamenal physical limit of ligand sensing 
\cite{int1} led us to
 conclude that receptor diffusion,
by removing bound ligand from a region of interest, may reduce the local overcounting  of  ligand molecules,  therefore potentially increasing the local accuracy of sensing. In this paper  we  set out to investigate this possibility 
by constructing  a mathematical model, that contains all the  necessary ingredients to describe the role of receptor diffusion in ligand sensing.
We have considered a receptor, that can bind and unbind ligand molecules as well as diffuse on a 2D membrane. Ligand is allowed to  diffuse in 3D space.  Using this model we derived the fluctuations in receptor occupancy in the area of interest via  a generalization  of the FDT  with the introduction of an effective temperature \cite{int1}.
 This ultimately allowed us  to derive  an equation for the accuracy of sensing.

We applied our model to the biologically relevant case of  glutamate receptors, which are  responsible for transmitting action potentials at neural synapses.
We found that for AMPAR and NMDAR,
our model shows
that the lateral diffusion of  receptors increases their accuracy of sensing and hence may allow synapses to count the number of released vesicles. However, the local occupancy fluctuations are decreased only for AMPAR; for NMDAR these fluctuations are  increased. This difference is due to the smaller unbinding rate of the latter receptor. Consequently, the 
accuracy of sensing  for  NMDAR is  an order of magnitude smaller than for  AMPAR.
It is important to remark that these results are consistent with the different roles  these  receptors play. In fact, the transmission of an excitatory potential through a synapse occurs in several steps. Initially, it is AMPAR which senses glutamate and, by  letting in ions, starts a small depolarization of the membrane. This  in turn allows  NMDAR to open its channel and start an even bigger influx of ions (especially Ca$^+$). This stimulates the production of more AMPARs and so increases the strength and sensitivity of the synapses.
Sensitivity to glutamate is important for AMPAR, while the roles of
NMDAR are coincidence detection and  amplification, important for long-term potentiation and memory.

 In order to derive the accuracy of sensing for diffusing receptors, we made a number of symplifying assumptions. We neglected fluctuations in the total receptor density and vesicle size,  assumed to be constant \cite{franks, aldo}.  Furthermore, we assumed
a constant ligand concentration  in  the hotspot area $\Delta A$  during the receptor measurement time, while in reality the  concentration profile will broaden out. Additionally, we made the simplifying assumption that multiple vesicles are released in the same spot, so that released glutamate concentrations add up, and that other effects such as 
spillover from neighboring cells can be neglected
\cite{nsites1,nsites2}. 
Finally, we introduced the
  effective temperature $T_e$  to generalize the FDT to non-equilibrum processes \cite{int1,cuglia,teo,crisanti,luca}}. The approximate nature of $T_e$, for which we neglected any potential time or frequency dependence,  limits the quantitative validity of our model to small deviations from equilibrium. In fact, for large receptor diffusion coefficients, the occupancy fluctuations and hence the accuracy of sensing become unphysical. 
  Nevertheless, the mapping of receptor diffusion to an effective internalization process, already studied in Ref. \cite{int1},
 provides confidence in our method.

In conclusion, in this paper we highlighted the role of diffusion
in increasing the accuracy of sensing. This might be important for synaptic counting   of glutamate vesicles. This hypothesis can 
be tested experimentally by  reducing the mobility of receptors. This can be achieved by cross-linking or addition of cholesterol. Alternatively, the glutamate diffusion costant can be reduced by adding dextran. Such perturbations
should lead to a broadening of the EPSC peaks, although signaling may be affected  as well.
Similarly to AMPAR \cite{renner}, an increase in receptor diffusion upon ligand binding was recently also observed in the pseudopods of migrating {\it Dictyostelium discoideum} cells \cite{keij}. This phenomenon was attributed to signal amplification. However, since cells may sense locally in pseudopods, at least in shallow gradients \cite{andrew}, receptor diffusion may also  be responsible for increasing the accuracy of sensing. 
%
%
 
%

\acknowledgments

We would like to thank Aldo Faisal and Peter Jonas for helpful comments.
We acknowledge financial support from
Biotechnological and Biological Sciences Research Council grant BB/G000131/1
and the Centre for Integrated Systems Biology at Imperial College.

\appendix

\section{Calculation of $\Sigma_1(\omega)$}
We devote this Appendix to the calculation of the integral $\Sigma_1(\omega)$ in Eq. (\ref{sigma1Int}). 
$\Sigma_1(\omega)$  is given by:
\begin{align}
\Sigma_1(\omega)=\int \frac{d \vec{q}}{(2 \pi)^2} \frac{1/D_2}{ q^2 +\frac{k_-}{D_2} - \frac{i\omega}{D_2}}=\int_0^{\infty} \frac{d q}{2 \pi} \frac{q/D2}{ q^2 +\frac{k_-}{D_2} - \frac{i\omega}{D_2} },
\end{align}
which is divergent. Hence, we introduce a cut-off $\Lambda \sim 1/s$ to account for the finite dimension  $s$ of the receptor.
This procedure leads to the final result:
\begin{align}
\Sigma_1(\omega)=\int_0^{\Lambda} \frac{d q}{2 \pi} \frac{q/D_2}{q^2 +\frac{k_-}{D_2} -\frac{i \omega }{D_2}}
=\frac{
\log\left(1+\frac{D_2 \Lambda^2}{k_- -i \omega}\right)}{4 \pi D_2}.
\end{align}

\section{Calculation of $\Sigma_2(\omega)$}
In this appendix, we calculate $\Sigma_2(\omega)$ in Eq. (\ref{sigma2Int}).
Without loss of generality, we set  $(x_0,y_0,z_0)=(0,0,0)$ and obtain:
\begin{flalign}
\nn &\Sigma_2(\omega)=\int_0^{\infty} \frac{dq_{\perp}}{2\pi}\int_0^{\infty} \frac{q dq}{2 \pi }\frac{i \omega/D_2 - q^2}{(q^2+q_{\perp}^2) - \frac{i\omega}{D_3}} \frac{1/D_3}{q^2 +\frac{k_-}{D_2} - \frac{i\omega}{D_2} }\\
\nn &=\int_0^{\infty} \frac{dq_{\perp}}{2\pi}\int_0^{\infty} \frac{ dq}{2 \pi}\frac{q/D_3}{(q^2+q_{\perp}^2)- \frac{i\omega}{D_3}}\left(\frac{k_-/D_2}{q^2 +\frac{k_-}{D_2} - \frac{i\omega}{D_2} }-1\right)\\
& =I_1(\omega)-I_2(\omega),
\end{flalign}

where

\begin{flalign}
\nn&I_1(\omega)=
 \int_0^{\infty} \frac{dq_{\perp}}{2\pi}\int_0^{\infty} \frac{q dq}{2 \pi}\frac{1/D_3}{(q^2+q_{\perp}^2)- \frac{i\omega}{D_3}}\frac{k_-/D_2}{q^2 +\frac{k_-}{D_2} - \frac{i\omega}{D_2} }\\
&= \int_0^{\Lambda} \frac{q dq}{8 \pi} \frac{i/D_3}{\sqrt{i \omega/D_3 - q^2} }\cdot\frac{k_-/D_2}{ q^2  +\frac{k_-}{D_2} -i \frac{\omega}{D_2} }\\
\nn &=\frac{k_-/(8 \pi D_3)}{\sqrt{\frac{k_-}{D_2} -i \omega(\frac{1}{D_2}-\frac{1}{D_3})}}\left[\arctan\left(\sqrt{\frac{\Lambda^2-i \omega/D_3}{\frac{k_-}{D_2} -i \omega (\frac{1}{D_2}-\frac{1}{D_3})}}\right) \right.\\
\nn &\left. -\arctan\left(\sqrt{\frac{-i \omega/D_3}{\frac{k_-}{D_2} -i \omega (\frac{1}{D_2}-\frac{1}{D_3})}}\right)\right]
\end{flalign}
and
\begin{flalign}
&I_2(\omega)=\int_0^{\infty} \frac{dq_{\perp}}{2\pi}\int_0^{\infty} \frac{q dq}{2 \pi}\frac{1}{-i \omega+D_3(q^2+q_{\perp}^2)}\\
\nn &=\frac{1}{D_3}\int_0^{\Lambda}\frac{q dq}{8 \pi}\frac{i}{\sqrt{-q^2+i \omega/D_3}}\\
\nn &= \frac{1}{8 \pi D_3}\left(\sqrt{- \frac{i\omega}{D_3}}-\sqrt{\Lambda^2- \frac{i\omega}{D_3}}\right)
\end{flalign} 
using  a cut off to account for the finite size of the receptor.
Taken together, the integral $\Sigma_2(\omega)$ is given by:
\begin{flalign}
\nn&\Sigma_2(\omega)=I_1(\omega)-I_2(\omega)=\left\lbrace\left(\sqrt{\Lambda^2- \frac{i\omega}{D_3}} -\sqrt{- \frac{i\omega}{D_3}}\right)+ \right.\\
\nn &\left. +\frac{k_-/ D_2}{\sqrt{\frac{k_-}{D_2} -i\omega (\frac{1}{D_2}-\frac{1}{D_3})}}\left[\arctan\left(\sqrt{\frac{ \Lambda^2-i \omega}{\frac{k_-}{D_2} -i \omega (\frac{1}{D_2}-\frac{1}{D_3})}}\right) \right.\right.\\
& \left. \left. -\arctan\left(\sqrt{\frac{-i \omega}{\frac{k_-}{D_2} -i \omega (\frac{1}{D_2}-\frac{1}{D_3})}}
\right)\right] \right\rbrace \frac{1}{8 \pi D_3}.
\end{flalign}

\end{document}